\documentclass[11pt,a4paper]{article}

\usepackage{amsmath, amsthm}            
\usepackage{amssymb}
\newcommand\eurfamily{\usefont{U}{eur}{m}{n}}   
\DeclareTextFontCommand{\texteur}{\eurfamily}
\usepackage{amscd}              
\usepackage{pslatex}
\usepackage[dvips]{epsfig}     

\allowdisplaybreaks

\renewcommand\smallskip{\vskip\smallskipamount}
\renewcommand\medskip{\vskip\medskipamount}
\renewcommand\bigskip{\vskip\bigskipamount}

\newcounter{bean}

\DeclareMathAlphabet{\BMi}{OT1}{cmm}{b}{it}     

\newcommand{\Z}{\mathbb{Z}}

\newcommand{\K}{\mathbb{K}}

\newcommand{\htensor}{\widehat{\otimes}}   

\newcounter{numb}

\theoremstyle{remark}

\theoremstyle{plain}

\theoremstyle{definition}

\begin{document}
\parindent=0pt

\begin{center}
{\Large\bf Formal Rigidity of the Witt and Virasoro Algebra}
\medskip

{\large Alice Fialowski\footnote{The research of the author was supported by
OTKA grant K77757.}}
\end{center}
\bigskip

{\sc Abstract} The formal rigidity of the Witt and Virasoro algebras
was first established by the author in \cite{F}. The proof was based on some earlier results
 of the author and Goncharowa, and was not presented there. In this
paper we give an elementary proof of these facts.
\bigskip
\begin{center}
{\bf I. Preliminaries}
\end{center}
\medskip
Consider the complexification $\sc{W}$ of the Lie algebra of polynomial
vector fields on the circle:
$$
e_k \to e^{ik\varphi}\, \frac{d}{d\varphi},
$$
where $\varphi$ is the angular parameter. The bracket operation in this
Lie algebra is 
$$
[e_n,e_m]=(m-n)e_{n+m}.
$$
The Lie algebra $\sc{W}$ is
called the \emph{Witt algebra}, and was first defined by E. Cartan
\cite{Ca}. The Lie algebra $\sc{W}$ is infinite dimensional and graded
with $\deg{e_n}=n$. It is well-known \cite{GF} that $\sc{W}$
has a unique nontrivial one-dimensional central extension. It is
generated by $e_n (n \in\Z)$ and the central element $c$, and its
bracket operation is defined by
$$
[e_n,e_m]=(m-n)e_{n+m} + 1/12(m^3-m)\delta_{n,-m}c,\quad [e_n,c]=0.
$$

The
extended Lie algebra $\sc{Vir}$ is called the \emph{Virasoro algebra}. 
It was firts invented for characteristic $0$ by Gelfand and Fuchs in \cite{GF}.

In this paper we give an elementary proof of the formal rigidity of
these algebras.
\bigskip
{\sc Formal deformations}
\medskip
Let $L$ be a Lie algebra, $A$ be a local finite dimensional  algebra over
 a field $\K$. 
\medskip
{\bf Definition. }(\text{see \cite{F1, F2}}) \emph{A formal deformation $L_A$ of $L$ parametrized by a
local finite dimensional algebra $A$ is a Lie algebra structure over $A$ on
$A\otimes_\K \, L$ such that the Lie algebra structure on}
$$
L = (L_A) \otimes_A\, \K = (L\otimes  A)\otimes_A \, \K
$$
\emph{is the given one on $L$}.
\medskip
Two deformations, $L_A$ and $L_A'$, parameterized by $A$ are called
\emph{equivalent}, if there exists a Lie algebra isomorphism over $A$
of $L_A$ on $L_A'$, inducing the identity of $L_A \otimes_A \, \K=L$ on 
$L_A' \otimes_A\, \K = L$. A deformation is \emph{trivial} if $A \otimes
L$ carries the trivially extended Lie structure.

Let now $A$ be a complete local algebra over $\K$, so
 $A= \lim\limits_{n \to\infty}(A/m^n)$,
 where $m$ is the maximal ideal of $A$ and we assume that $A/m \cong
 \K$.
 \medskip
 A \emph{formal deformation of $L$ with base $A$} is a Lie algebra structure
 over $A = \overleftarrow{\lim\limits_{n\to \infty}}(A/m^n)$ on the completed 
 tensor product $A \textstyle{\htensor}_{\K} L = \lim\limits_{n
 \to\infty}((A/m^n) \textstyle{\htensor} L)$ such that
$$
\epsilon\textstyle{\htensor}
\text{id}: A \textstyle{\htensor}L \to \K\otimes L = L
$$
is a Lie algebra homomorphism.

There is an analogous definition for equivalence of deformations
parameterized by a complete local algebra. 
 
 \medskip
 
{\sc Rigidity}.
\smallskip
Intuitively, rigidity of a Lie algebra means that we cannot deform the
Lie algebra.
 
{\bf Definition}.
A Lie algebra $L$ is formally \emph{rigid}, if and only if every formal
deformation of it is equivalent to the trivial deformation.

{\bf Proposition}. The elements of the cohomology space $H^2(L;L)$ correspond
bijectively to the nonequivalent infinitesimal deformations. 

{\bf Corollary}. The condition $H^2(L;L)=0$ is sufficient for $L$ to be
rigid. 
\smallskip
For details see the book of Fuchs, \cite{Fu}.
\bigskip
 
\begin{center}
{\bf II. Rigidity of $W$ and $Vir$}
\end{center}
\bigskip
{\bf Theorem}. The Witt and Virasoro algebra are formally rigid. 
\medskip
{\bf Proof}. We will prove that $H^2(W;W)=0$. It follows than that
$H^2(Vir;Vir)=0$ as well. (Another, more complicated proof is in
\cite{FSch}.)
\smallskip

We will present the proof in 9 steps. For the basics of Lie algebra
cohomology see the book \cite{Fu}.
\smallskip

{\bf 1.} First assume that the 2-cochain ;$c$ has weight $d$: 
$c\in C^2_d(W;W)$ which means that
$c(e_i,e_j)=c_{i,j}e_{i+j+d}$. We will prove that if $d\ne0$, and
$\delta c=0$, then $c=\delta b$ where
$b(e_i)=\displaystyle{\frac{c(e_i,e_0)}{ d}}$ (in particular, $b(e_0)=0$).
Indeed, let $c'=c-\delta b$. Then,
first,
\begin{align*}
c'(e_i,e_0)&=c(e_i,e_0)-\delta b(e_i,e_0)\\
&=d\cdot b(e_i)-b([e_i,e_0])+[e_i,b(e_0)]-[e_0,b(e_i)]\\
&=d\cdot b(e_i)+i\cdot b(e_i)-(i+d)b(e_i)=0.
\end{align*}

Second, since $\delta c'=0$, we have
\begin{align*}
\delta c'(e_i,e_j,e_0)&=(j-i)c'(e_{i+j},e_0)+ic'(e_i,e_j)-jc'(e_j,e_i)\\
&=-[e_i,c'(e_j,e_0)]+[e_j,c'(e_i,e_0]]-[e_0,c'(e_i,e_j)]\\
&=(i+j-(i+j+d))c'(e_i,e_j)=-d\cdot c'(e_i,e_j)=0.
\end{align*}
Hence, $c'=0$. (See also Theorem 1.5.2 in \cite{Fu}.)  

This shows that $H^2(W;W)=H^2_0(W;W)$. Indeed, if
$c\in C^2(W;W)$ and $\delta c=0$, then $c-\delta b\in C^2_0(W;W)$
where $b$ is the following: if $c(e_i,e_j)=\sum\limits_d
c_{i,j;d}e_{i+j+d}$, then
$b(e_i)=\sum\limits_{d\ne0}\dfrac{c_{i,0,d}}{d}e_{i+d}$.

\noindent Thus, from now on, we consider a cocycle\ 
$c\in C^2_0(W;W),\ c(e_i,e_j)=c_{i,j}e_{i+j},\ \ \delta c=0$. Denote
$c$ by $\{c_{i,j}\}$.
\bigskip

{\bf2.} First, let us explore the possibility of adding $\delta b,\
b\in C^1_0(W;W)$, such that $b(e_i)=b_ie_i$, to $c$.  Since $\{b_i=i\}$
is a coboundary and hence a cocycle, we can assume $b_1=0$.
\smallskip

{\sc Proposition.} \emph{Varying $b$, we can achieve}
$$
c_{i,1}=0 \text{\quad for all }\, i, \text{\quad   and \ }
c_{-2,2}=0.\eqno(1)
$$
\emph{Moreover, this is all that can be done by adding $\delta b$ to $c$.}
\smallskip
{\sc Proof.} Indeed, 
$$
\delta b(e_{-2},e_2)=4(b_0-b_2-b_{-2})e_0,\quad
\delta b(e_i,e_1)=(1-i)e_{i+1}.\eqno(2)
$$
The second of these formulas gives
\begin{align*}
\delta b(e_0,e_1)=&\hskip.40in-\, b_0\,e_1
     & \delta b(e_2,e_1)=&-\ (b_3-b_2)e_3\\
\delta b(e_{-1},e_1)=&2(b_0-b_{-1})\ e_0    \tag{3}
     & \delta b(e_3,e_1)=&-2(b_4-b_3)e_4 \\ 
\delta b(e_{-2},e_1)=&3(b_{-1}-b_{-2})e_{-1} & \delta
b(e_4,e_1)=&-3(b_5-b_4)e_3\\
\dots&\dots &\dots&\dots
\end{align*}
First, using the first column of (3) and choosing appropriate
$b_0$, $b_{-1}$, $b_{-2},\dots$, we can make $\delta b(e_i,e_1)=c(e_i,e_1)$
for $i\le0$. Second, using the left formula (2) and choosing an
appropriate $b_2$ we can make $\delta b(e_{-2},e_2)=c(e_{-2},e_2)$.
Third, using the second column of (3) and choosing appropriate $b_3,
b_4, b_5,\dots$ we can make $\delta b(e_i,e_1)=c(e_i,e_1)$ for
$i\ge2$. 
\smallskip

Notice that this construction uses a unique choice of all $b_i$
(excluding $b_1$ which is 0).
\smallskip

{\sc Corollary.} {\it Every cohomology class in $H^2(W;W)$ is
represented by a unique cocycle $\{c_{i,j}\}$ satisfying relations}
(1).
\bigskip

{\bf3. The cocycle condition.} It remains to prove that  if
$c=\{c_{i,j}\}$ as above satisfies the condition $\delta c=0$, then
$c=0$. Remark that the condition $\delta c(e_i,e_j,e_k)=0$ takes the
form
\begin{align*}
(j-i)&c_{i+j,k}+(k-j)c_{j+k,i}+(i-k)c_{k+i,j}\\   \tag{4}
   +&(j-i+k)c_{k,j}+(j-i-k)c_{k,i}-(i+j-k)c_{i,j}=0. 
\end{align*}

{\bf4. Set $\bf k=1$.} Because of the results of Section 2, three of
the six terms in (4) vanish, and what remains
is$$(j-1)c_{i,j+1}+(i-1)c_{i+1,j}-(i+j-1)c_{i,j}=0.\eqno(5)$$We will use abbreviated notation $a_j=c_{2,j}$. We already know that $a_{-2}=a_1=a_2=0$.\smallskip
\bigskip

{\bf 5.} {\sc Proposition.} {\it If $i\le0$, then}$$c_{ij}=0\ \ {\rm
  for }\ j\le1,\ {\rm and}\ c_{ij}=(i-1)a_0\ \ {\rm for }\ j\ge-i+2.$$

{\sc Proof.} We proceed by induction with respect to $-i$. First, take $i=0$.
 By the Proposition in Section 2, 
equation (5) loses one more term, and becomes
$(j-1)(c_{0,j+1}-c_{0,j})=0$. This
gives
$$
\dots=c_{0,-2}=c_{0,-1}=c_{0,0}=c_{0,1};\
c_{0,2}=c_{0,3}=c_{0,4}=\dots.
$$
Since $c_{0,2}=-b_0$ and $c_{0,1}=0$, this is precisely what we need in
this case.
\smallskip
Assume now that $i<0$ and that the formulas in our Proposition hold for
$c_{i+1,j}.$ Then for $j\le0$ the middle term in (5) vanishes and we
have $(i+j-1)c_{i,j}=(j-1)c_{i,j+1}$, and, since $c_{i,1}=0$, this
formula with $j=0,-1,-2,\dots$ successively gives
$c_{i,0}=0,c_{i,-1}=0,c_{i,-2}=0,\dots$. For $j\ge-i+1$, formula
(5) becomes
$$
(j-1)c_{i,j+1}=(i+j-1)c_{i,j}-i(i-1)a_0.
$$
If $j=-i+1$, this becomes $-ic_{i,-1+2}=-i(i-1)a_0$ which implies
$c_{i,-i+2}=(i-1)a_0$.
If we already know that $c_{i,j}=(i-1)a_0$, then
$$(
j-1)c_{i,j+1}=((i+j-1)(i-1)-i(i-1))a_0=(j-1)(i-1)a_0.
$$
Consequently, $c_{i,j+1}-(i-1)a_0$.
\smallskip

All that we know now about the cocycle $\{c_{i,j}\}$ is presented in
the following table:
\medskip


\begin{center}
\vbox{\offinterlineskip
\hrule
\halign{&\vrule#&\hskip4pt\hfil#\hfil\hskip4pt\strut\cr
height2pt&\omit&&\omit&&\omit&&\omit&&\omit&&\omit&&\omit&&\omit&&\omit&&\omit&&\omit&\cr
& &&$j=-4$&&$j=-3$&&$j=-2$&&$j=-1$&&$j=0$&&$j=1$&&$j=2$&&$j=3$&&$j=4$&&$j=5$&\cr
height2pt&\omit&&\omit&&\omit&&\omit&&\omit&&\omit&&\omit&&\omit&&\omit&&\omit&&\omit&\cr
\noalign{\hrule}
height2pt&\omit&&\omit&&\omit&&\omit&&\omit&&\omit&&\omit&&\omit&&\omit&&\omit&&\omit&\cr
&$i=5$&& &&$4a_0$ &&$3a_0$&&$2a_0$&&$a_0$&&0&&$-a_5$&& && &&\bf0&\cr
height2pt&\omit&&\omit&&\omit&&\omit&&\omit&&\omit&&\omit&&\omit&&\omit&&\omit&&\omit&\cr
\noalign{\hrule}
height2pt&\omit&&\omit&&\omit&&\omit&&\omit&&\omit&&\omit&&\omit&&\omit&&\omit&&\omit&\cr
&$i=4$&& && &&$3a_0$&&$2a_0$&&$a_0$&&0&&$-a_4$&& &&\bf0&& &\cr
height2pt&\omit&&\omit&&\omit&&\omit&&\omit&&\omit&&\omit&&\omit&&\omit&&\omit&&\omit&\cr
\noalign{\hrule}
height2pt&\omit&&\omit&&\omit&&\omit&&\omit&&\omit&&\omit&&\omit&&\omit&&\omit&&\omit&\cr
&$i=3$&& && && &&$2a_0$&&$a_0$&&0&&$-a_3$&&\bf0&& && &\cr
height2pt&\omit&&\omit&&\omit&&\omit&&\omit&&\omit&&\omit&&\omit&&\omit&&\omit&&\omit&\cr
\noalign{\hrule}
height2pt&\omit&&\omit&&\omit&&\omit&&\omit&&\omit&&\omit&&\omit&&\omit&&\omit&&\omit&\cr
&$i=2$&&$a_{-4}$&&$a_{-3}$&&0&&$a_{-1}$&&$a_0$&&0&&\bf0&&$a_3$&&$a_4$&&$a_5$&\cr
height2pt&\omit&&\omit&&\omit&&\omit&&\omit&&\omit&&\omit&&\omit&&\omit&&\omit&&\omit&\cr
\noalign{\hrule}
height2pt&\omit&&\omit&&\omit&&\omit&&\omit&&\omit&&\omit&&\omit&&\omit&&\omit&&\omit&\cr
&$i=1$&&0&&0&&0&&0&&0&&\bf0&&0&&0&&0&&0&\cr
height2pt&\omit&&\omit&&\omit&&\omit&&\omit&&\omit&&\omit&&\omit&&\omit&&\omit&&\omit&\cr
\noalign{\hrule}
height2pt&\omit&&\omit&&\omit&&\omit&&\omit&&\omit&&\omit&&\omit&&\omit&&\omit&&\omit&\cr
&$i=0$&&0&&0&&0&&0&&\bf0&&0&&$-a_0$&&$-a_0$&&$-a_0$&&$-a_0$&\cr
height2pt&\omit&&\omit&&\omit&&\omit&&\omit&&\omit&&\omit&&\omit&&\omit&&\omit&&\omit&\cr
\noalign{\hrule}
height2pt&\omit&&\omit&&\omit&&\omit&&\omit&&\omit&&\omit&&\omit&&\omit&&\omit&&\omit&\cr
&$i=-1$&&0&&0&&0&&\bf0&&0&&0&&$-a_{-1}$&&$-2a_0$&&$-2a_0$&&$-2a_0$&\cr
height2pt&\omit&&\omit&&\omit&&\omit&&\omit&&\omit&&\omit&&\omit&&\omit&&\omit&&\omit&\cr
\noalign{\hrule}
height2pt&\omit&&\omit&&\omit&&\omit&&\omit&&\omit&&\omit&&\omit&&\omit&&\omit&&\omit&\cr
&$i=-2$&&0&&0&&\bf0&&0&&0&&0&&0&& &&$-3a_0$&&$-3a_0$&\cr
height2pt&\omit&&\omit&&\omit&&\omit&&\omit&&\omit&&\omit&&\omit&&\omit&&\omit&&\omit&\cr
\noalign{\hrule}
height2pt&\omit&&\omit&&\omit&&\omit&&\omit&&\omit&&\omit&&\omit&&\omit&&\omit&&\omit&\cr
&$i=-3$&&0&&\bf0&&0&&0&&0&&0&&$-a_{-3}$&& && &&$-4a_0$&\cr
height2pt&\omit&&\omit&&\omit&&\omit&&\omit&&\omit&&\omit&&\omit&&\omit&&\omit&&\omit&\cr
\noalign{\hrule}
height2pt&\omit&&\omit&&\omit&&\omit&&\omit&&\omit&&\omit&&\omit&&\omit&&\omit&&\omit&\cr
&$i=-4$&&\bf0&&0&&0&&0&&0&&0&&$-a_{-4}$&& && && &\cr
height2pt&\omit&&\omit&&\omit&&\omit&&\omit&&\omit&&\omit&&\omit&&\omit&&\omit&&\omit&\cr
\noalign{\hrule}
}
\hrule}
\end{center}

\smallskip
Nothing is known, so far, about the values in the empty cells. 
\bigbreak

{\bf6.} Formula (5) with $i=2,3,\dots$ gives an expression of
$c_{i,j},\ i\ge3$ in terms of $a_k$; thus, we can fill in the remaining
cells in the last table. In
particular,
\begin{align*}
c_{3,j}&=(j+1)a_j-(j-1)a_{j+1};\\
c_{4,j}&=\frac{(j+1)(j+2)}{2}a_j-(j-1)(j+2)a_{j+1}+\frac{j(j-1)}{2}a_{j+2};\\
c_{5,j}&=\frac{(j+1)(j+2)(j+3)}{6}a_j -
\frac{(j-1)(j+2)(j+3)}{2}a_{j+1} \\
&\qquad\qquad +\frac{(j-1)j(j+3)}{2}a_{j+2} -\frac{(j-1)j(j+1)}{6}a_{j+3};
\end{align*}
etc. (It is not hard to write a general formula, but we will not need
it.) We see that all the $c_{ij}$-s are expressed as linear combinations of
$a_k$, so all we need is to prove that all $a_k$ are zero. Notice also
that the equalities $c_{i,i}=0$ provide some new relations between
$a_k$:
\begin{align*}
&c_{3,3}=0\ \Rightarrow\  2a_3-a_4=0\ \Rightarrow\ 
a_4=2a_3;\\ &c_{4,4}=0\ \Rightarrow\ 5a_4-6a_5+2a_6=0\ \Rightarrow\ 
	a_6=3a_5-5a_3;\\ 
&c_{5,5}=0\ \Rightarrow\ 14a_5-28a_6+20a_7-5a_8\ \Rightarrow\ 
	a_8=4a_7-14a_5+28a_3;\\ 
&c_{6,6}=0\ \Rightarrow\ \dots\ \Rightarrow\ 
	a_{10}=5a_9-30a_7+117a_5-255a_3;
\end{align*}
etc. This can be continued to provide good-looking formula expressing
$a_{2k},\, k\ge2$ in terms of $a_{2\ell-1},\, \ell\ge2$; however, in 
our further computations we will only use the first of the relations
above ($a_4=2a_3$).
\bigskip

{\bf7. Put $\bf k=2$.} Formula (5)
becomes
\begin{align*}
\hskip 3cm (i+j-2)&c_{i,j}-(i-2)c_{i+2,j}-(j-2)c_{i,j+2} \\
  &=(i-j)a_{i+j}-(i-j+2)a_i-(i-j-2)a_j;\hskip 1.5cm(6)
\end{align*}
this will be a source of relations between $a_k$ which will kill all of
them.
\bigskip

{\bf8. \ Let $\bf i=-2$.} Formula (5)
becomes
$$
(j-4)c_{-2,j}+4c_{0,j}-(j-2)c_{-2,j+2}=-(j+2)a_{j-2}+(j+4)a_j.\eqno(7)
$$
We know that
$$
c_{0,j}=\begin{cases}
0,&\text{if $j\le1$},\\
 -a_0,&\text{if $j\ge2$};\end{cases}
\qquad
c_{-2,j}=\begin{cases}0,&\text{if $j\le2$},\\ 
-3a_{-1},&\text{if $j=3$},\\
 -3a_0,& \text{if $j\ge4$}\end{cases}.
$$
Note that we used a formula from Section 6:
$$
c_{-2,3}=-c_{3,-2}=-[(-2+1)a_{-2}-(-2-1)a_{-1}]=-3a_{-1}.
$$
Thus, for $j\le0$, the left hand side of (7) is zero, and we have
$(j+4)a_j=(j+2)a_{j-2}$. Thus, 
$$
a_0=0,\quad 0=a_{-6}=a_{-8}=a_{-10}=\dots
$$
and
$$
3a_{-1}=a_{-3}=-a_{-5}=-3a_{-7}=-5a_{-9}=-7a_{-11}=\dots
$$
Since $a_0=0$, the left hand side of formula (7) is zero also for
$j\ge4$. Thus,
$$
 (j+4)a_j=(j+2)a_{j-2}\ \text{also for}\ j\ge4.
 $$
  Hence,
$$
0=a_2=a_4=a_6=\dots,\qquad a_3=a_5=a_7=\dots.
$$
But $a_4=2a_3$ (see the end of Section 6), so actually, $a_k=0$ for all
$k>0$.  Thus, there remain two unknown values: $a_{-4}$ and $a_k$ for
any one odd negative $k$.
\bigskip

{\bf9. Last Step.} Consider formula (6) for $i=-3$:
\begin{align*}
(j-5)c_{-3,j}&+5c_{-1,j+2}-(j-2)c_{-3,j+2}\\
  &=-(j+3)a_{j-2}+(j+1)a_{-3}+(j+5)a_j.
\end{align*}
If $j=-6$ (or $-8,-10,$ etc.), we get $a_{-3}=0$; if $j=-4$, we get
$a_{-4}=0$. This completes the calculation.

\bigskip
\parindent 50pt \obeylines
     \vbox{{\sc A. Fialowski}
 Institute of Mathematics
 E\"otv\"os Lor\'and University
 P\'azm\'any P\'eter s\'et\'any 1/C
 H-1117 Budapest, Hungary
 {\tt fialowsk@cs.elte.hu}}

\end{document}